\documentclass[%
reprint,
showpacs,preprintnumbers,
longbibliography,
amsmath,amssymb,
aps,
prb,
floatfix,
]{revtex4-1}
\usepackage[dvipdfmx]{graphicx}
\usepackage[dvipdfmx]{color}
\usepackage{dcolumn}
\usepackage{tabularx}
\usepackage{booktabs}
\usepackage{bm}
\usepackage{float}
\usepackage{braket}
\usepackage{here}
\usepackage{comment}
\newcommand{\argmin}{\mathop{\rm argmin}\limits}

\begin{document}
\title{Prediction of the Curie temperature considering the dependence of the phonon free energy on magnetic states}
\author{Tomonori Tanaka}
 \email{tanaka.t.bj@m.titech.ac.jp}
\author{Yoshihiro Gohda}
 \email{gohda.y.ab@m.titech.ac.jp}
\affiliation{Department of Materials Science and Engineering, Tokyo Institute of Technology, Yokohama 226-8502, Japan}
\date{\today}
\begin{abstract}
Prediction of the Curie temperature is of significant importance for the design of ferromagnetic materials.
Even though the Curie temperature has been estimated using the Heisenberg model, magnetic exchange coupling parameters
widely used is thus far based on first-principles calculations at zero temperature.
In the explicit consideration of temperature effects, it is important to minimise the total free energy, because the magnetic and phonon free energies correlate with each other.
Here, we propose a first-principles thermodynamic approach to minimise the total free energy considering
both the influences of magnetism on phonons and the feedback effect from phonons to magnetism.
By applying our scheme to bcc Fe, we find a significant reduction of the Curie temperature due to the feedback effect.
This result inevitably enforces us to change our convention as follows: we should use exchange coupling constants for 
the disordered local moment state, not for the ferromagnetic state, in the prediction of the Curie temperature.
Our results not only change the fundamental understanding of finite-temperature magnetism but also provide
a general framework to predict the Curie temperature more accurately.
\end{abstract}
\pacs{}
\maketitle

The Curie temperature ($T_{\rm C}$) is one of the essential properties of ferromagnetic materials
because it characterises their applicability and performance\cite{mag_apl, coey}.
The method of predicting $T_{\rm C}$ is, therefore, important
not only for a fundamental understanding of ferromagnetic materials but also for the material design for applications.
A typical technique for predicting $T_{\rm C}$ is a downfolding method from first-principles calculations to an effective lattice model as below:
Firstly, one derives exchange coupling constants ($J_{ij}$) by applying Green's function-based methods\cite{oguchi, mag_force} or
a frozen magnon approach\cite{halilov1, halilov2}.
Secondly, one builds an effective lattice model such as the Heisenberg model and assign $J_{ij}$ to the model.
Finally, one solves the model analytically or numerically and estimates $T_{\rm C}$.
This technique is applied to a broad range of materials,
such as 3$d$ transition metals\cite{mag_force,sakuma,pajda,takahashi,FeGa,realistic}
and rare-earth magnets\cite{Smtype,kashyap,turek,hcpGd,NdFeB,gong}.
Such many studies demonstrated that the prediction technique has some predictive accuracy.

Such a technique usually does not include temperature effects on magnetic interactions.
Moreover, temperature-induced interactions between magnetism and other excitations such as phonons sometimes
make the accurate prediction of $T_{\rm C}$ difficult.
At a high-temperature range around $T_{\rm C}$, there are two kinds of interaction between magnetism and phonons.
One is the effect of thermal atomic displacements on $J_{ij}$\cite{sabiryanov, yin_vib, ruban_vib}.
The change in $J_{ij}$ obviously modifies $T_{\rm C}$.
The other interaction is the effect of magnetic disordering on phonon frequencies.
Some ferromagnetic materials such as bcc Fe and $\rm{Pd_3Fe}$ shows phonon softening at elevated temperatures near $T_{\rm C}$\cite{softening_exp1,softening_exp2,pd3fe}.
Some research groups approached this phenomenon by different theoretical methods\cite{paramag,dmft1,dmft2,mag_ph,dmft3,tdep,dutta_phonons,pd3fe}
and achieved the same conclusion:
The phonon softening is due to magnetic disordering near $T_{\rm C}$.
Regarding the predictive accuracy of $T_{\rm C}$, the importance of the former interaction is easily understandable,
whereas the latter interaction does not apparently seem to be related to $T_{\rm C}$.
However, we will recognise the phonon softening due to magnetic disordering is closely related to $T_{\rm C}$ by standing a thermodynamic viewpoint.

Thermal equilibrium states at finite temperature correspond to the minimum of the total free energy at given conditions.
This is usually called as the minimum principle for the free energy.
Usual procedures to study finite-temperature magnetism is constructing a magnetic Hamiltonian and deriving thermodynamic quantities such as the magnetic energy and the magnetisation.
This series of procedures is equal to interpret that equilibrium magnetic quantities are determined through the magnetic free energy only.
This interpretation, however, collapses in the systems that exhibit the phonon softening due to magnetic disordering.
The phonon frequencies are directly related to the phonon free energy.
Thus the phonon softening due to magnetic disordering means that magnetic states affect the phonon free energy as well as the magnetic free energy.
As a result, equilibrium magnetic states should be determined through not only the magnetic free energy but also
the phonon free energy, according to the minimum principle for the free energy.
We call this effect of phonons on equilibrium magnetic states through the change of the phonon free energy as a thermodynamic feedback effect.
This feedback effect surely affects $T_{\rm C}$ as a consequence of the change of equilibrium magnetic states.
However, the significance of the feedback effect on $T_{\rm C}$ is unclear because the existence of the effect has been overlooked.

In this article, we propose a thermodynamic formulation to treat the feedback effect from phonons to magnetism.
The formulation results in a simple optimisation problem for the total free energy.
The ingredients to solve the problem
are evaluated by first-principles phonon calculations and Monte Carlo simulations based on the Heisenberg model.
By applying the formulation to bcc Fe, we demonstrate that $T_{\rm C}$ of bcc Fe significantly decreases by nearly 580 K.
This result proves the feedback effect is crucial for accurate prediction of $T_{\rm C}$.
We also discuss the relationship between the predictive accuracy of $T_{\rm C}$ and reference magnetic states in the derivation of $J_{ij}$.
Remarkably, we find a significant overestimation of $T_{\rm C}$ in a paramagnetic disordered local moment (DLM) state is rather a correct tendency.
Quantitative description of finite-temperature magnetism plays an important role in both basic and applied materials science.
Therefore, our results have an impact on the fundamental understanding of magnetism and materials design for ferromagnetic materials.

We organise the following part of this paper as below.
Firstly, we introduce a new thermodynamic formulation to treat the thermodynamic feedback effect.
Our formulation based on the minimum principle for the free energy is justified through the Legendre transformation and 
results in a simple optimisation problem.
Next, we evaluate the magnetic entropy and the phonon free energy of bcc Fe as functions of the magnetic energy.
These functions are needed to solve the optimisation problem.
Finally, we evaluate the equilibrium magnetic energy around $T_{\rm C}$ by solving the optimisation problem.
The shift of $T_{\rm C}$ of bcc Fe is estimated from the results of the equilibrium magnetic energy.

\section*{Thermodynamic formulation for magnetic materials}
In conventional thermodynamic approaches for magnetic materials, the phonon and magnetic contributions are assessed independently.
We start from this typical case for comparison with our formulation.
The fundamental relation is written as
\begin{align}
E_{\rm tot}(S_{\rm ph}, S_{\rm mag}) \approx E_{\rm ph}(S_{\rm ph}) + E_{\rm mag}(S_{\rm mag}),
\end{align}
where $E$ is the energy and $S$ is the entropy. The subscripts tot, ph and mag represent total, phonon and magnetic, respectively.
Here, we consider the Gibbs free energy,
\begin{align}
G(T, p, H) = E-TS+pV-\mu_0 MH,
\end{align}
where $T$ represents the temperature, $p$ the pressure, $V$ the volume, $M$ the magnetisation, $H$ the external magnetic field and $\mu_0$ the vacuum permeability.
In the following, we derive the formalism for $p=0$ and $H=0$, but the discussion remains unchanged for the case with finite external fields $p$ and $H$.
The Gibbs free energy $G$ is derived by applying the Legendre transformation.
\begin{align}
G&_{\rm tot}(T) = \underset{S_{\rm ph}, S_{\rm mag}}{\rm min}\{E_{\rm ph}(S_{\rm ph}) - TS_{\rm ph} \notag \\
&\ \ \ \ \ \ \ \ \ \ \ \ \ \ \ \ \ \ \ \ \ \ \ \ \ + E_{\rm mag}(S_{\rm mag}) - TS_{\rm mag}\}\\
 &= \underset{E_{\rm ph}, E_{\rm mag}}{\rm min}\{E_{\rm ph} - TS_{\rm ph}(E_{\rm ph}) \notag \\
&\ \ \ \ \ \ \ \ \ \ \ \ \ \ \ \ \ \ \ \ \ \ \ \ \ + E_{\rm mag} - TS_{\rm mag}(E_{\rm mag})\}\\
 &= \underset{E_{\rm ph}, E_{\rm mag}}{\rm min} \{G_{\rm ph} (T, E_{\rm ph})
 + G_{\rm mag}(T, E_{\rm mag})\}\\
&= G_{\rm ph}(T) +  G_{\rm mag}(T).
\end{align}
Here we used the one-to-one correspondence between energy and entropy for fixed other thermodynamic parameters such as $V$ and $M$.
The independent assessment in conventional approaches is based on this trivial formulation.
\begin{figure}
\includegraphics[width=\linewidth]{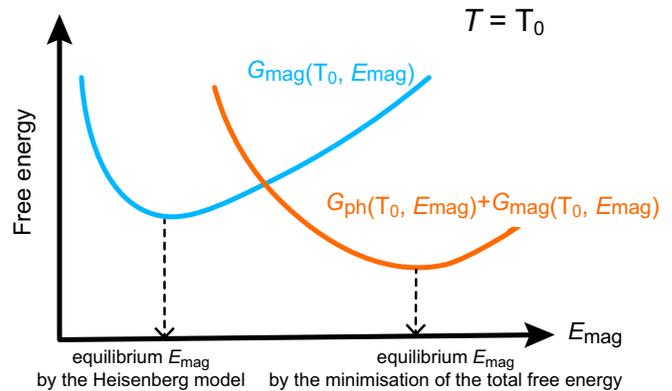} 
\caption{Schematic image of the free energy minimisation at a temperature ${\rm T}_{0}$.
Within a common framework using the Heisenberg model, the equilibrium magnetic energy ($E_{\rm mag}$) is corresponding to the minimum of
the magnetic free energy, $G_{\rm mag}$ (blue line).
On the other hand, the equilibrium magnetic energy in our scheme is corresponding to the minimum of the total free energy, $G_{\rm ph}+G_{\rm mag}$ (orange line).}
\label{minimisation}
\end{figure}
\begin{figure*}
\includegraphics[width=\linewidth]{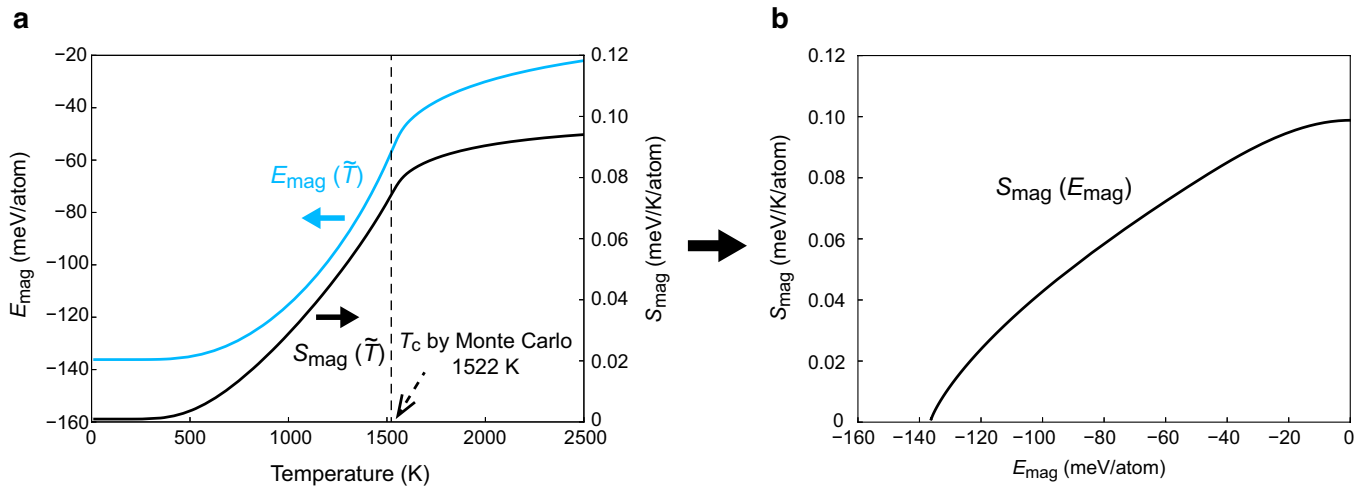} 
\caption{Thermodynamic quantities of bcc Fe obtained by the rescaled Monte Carlo method.
(a) Energy and entropy vs. temperature.
(b) Entropy vs. energy.
The theoretical Curie temperature $T_{\rm C}$ was identified from the peak of the specific heat.
}
\label{rMC_results}
\end{figure*}
Next, we incorporate the dependence of the phonon free energy on magnetic states.
We assume that the magnitude of the interaction between magnetic disordering and phonon frequencies can be written as thermodynamic quantities of the magnetic part.
K\"ormann {\it et al.} proposed a solid treatment with this assumption\cite{mag_ph}.
They treated the forces on each atom as a function of the magnetic energy.
As a result, the phonon frequencies, consequently the phonon free energy, have the dependence on the magnetic energy (see Methods).
Thermodynamically speaking, their treatment means the phonon energy depends not only on the phonon entropy but also on the magnetic entropy.
The fundamental relation thus can be written as
\begin{equation}
E_{\rm tot}(S_{\rm ph}, S_{\rm mag}) \approx E_{\rm ph}(S_{\rm ph}, S_{\rm mag}) + E_{\rm mag}(S_{\rm mag}).
\end{equation}
In principle, $E_{\rm mag}$ also has a dependence on $S_{\rm ph}$.
This dependence can be regarded as influences of thermal atomic displacements on $J_{ij}$\cite{ruban_vib}.
If we want to incorporate this effect into the thermodynamic formulation, we have to express the magnitude of the effect as a thermodynamic quantity such as $S_{\rm ph}$.
However, the correspondence between the thermal displacements and $S_{\rm ph}$ is not obvious.
We thus focus only the dependence of $E_{\rm ph}$ on $S_{\rm mag}$.

We apply the Legendre transformation as before.
\begin{align}
G&_{\rm tot}(T)  = \underset{S_{\rm ph}, S_{\rm mag}}{\rm min}\{E_{\rm ph}(S_{\rm ph}, S_{\rm mag}) - TS_{\rm ph} \notag \\
&\ \ \ \ \ \ \ \ \ \ \ \ \ \ \ \ \ \ \ \ \ \ \ \ \ + E_{\rm mag}(S_{\rm mag}) - TS_{\rm mag}\}\\
 &=  \underset{S_{\rm mag}}{\rm min}\{G_{\rm ph}(T, S_{\rm mag}) + E_{\rm mag}(S_{\rm mag}) - TS_{\rm mag}\}\\
 &= \underset{E_{\rm mag}}{\rm min}\{G_{\rm ph}(T, E_{\rm mag}) + E_{\rm mag} - TS_{\rm mag}(E_{\rm mag})\} \\
 &= \underset{E_{\rm mag}}{\rm min}\{G_{\rm ph}(T, E_{\rm mag}) + G_{\rm mag}(T, E_{\rm mag})\}.
\end{align}
Note that the entropy (or energy) and the temperature can be treated as independent variables during the minimisation procedure.
The thermodynamic relationship between the entropy and the temperature, such as $\partial G/\partial T=-S$,
holds after the minimisation, i.e. after the Legendre transformation.
The last expression is very intuitive from a thermodynamic viewpoint:
The equilibrium magnetic energy at a temperature $\rm T_0$ is determined to minimise the total free energy (Fig. \ref{minimisation}) as
\begin{align}
&\argmin_{E_{\rm mag}}\left[ G_{\rm ph}({\rm T_0}, E_{\rm mag}) + E_{\rm mag} - {\rm T_0}S_{\rm mag}(E_{\rm mag}) \right] \label{argmin}\\
=&\argmin_{E_{\rm mag}}\left[ G_{\rm ph}({\rm T_0}, E_{\rm mag}) + G_{\rm mag}({\rm T_0}, E_{\rm mag})\right].
\end{align}

\section*{evaluations of the magnetic entropy and the phonon free energy}
We demonstrate the significance of the dependence of the phonon free energy on magnetic states for bcc Fe as an example.
As a starting point, we evaluate the magnetic entropy and the phonon free energy depending on the magnetic energy
($S_{\rm mag}(E_{\rm mag})$ and $G_{\rm ph}(T, E_{\rm mag})$),
in order to solve the minimisation problem in equation (\ref{argmin}).

To obtain $S_{\rm mag}(E_{\rm mag})$, we carried out the rescaled Monte Carlo method\cite{rescaled} based on the Heisenberg model.
This method brings thermodynamic quantities derived from classical Monte Carlo simulations closer to those from quantum Monte Carlo simulations.
The exchange coupling constants ($J_{ij}$) in the Heisenberg model are derived from the paramagnetic disordered local moment (DLM) state\cite{oguchi,gyorffy} (see Methods).
The magnetic energy and entropy as functions of lattice-model temperature $\widetilde{T}$ are shown in Fig. \ref{rMC_results} (a).
The theoretical $T_{\rm C}$ (1522 K) is higher than the experimental value (1043 K).
Such overestimation has also been reported in previous studies\cite{oguchi,chana,ruban_vib} using the DLM state.
The overestimation has been recognised as a disadvantage of the DLM state, and it will be discussed later associated with our results.
Since this magnetic system does not show the first-order phase transition,
the one-to-one correspondence holds between not only $E_{\rm mag}$ and $S_{\rm mag}$ but also $\widetilde{T}$,
\begin{equation}
E_{\rm mag} \leftrightarrow \widetilde{T} \leftrightarrow S_{\rm mag}.
\end{equation}
We constructed the function $S_{\rm mag}(E_{\rm mag})$ (Fig. \ref{rMC_results} (b)) by using this relationship.

\begin{figure}
\includegraphics[width=\linewidth]{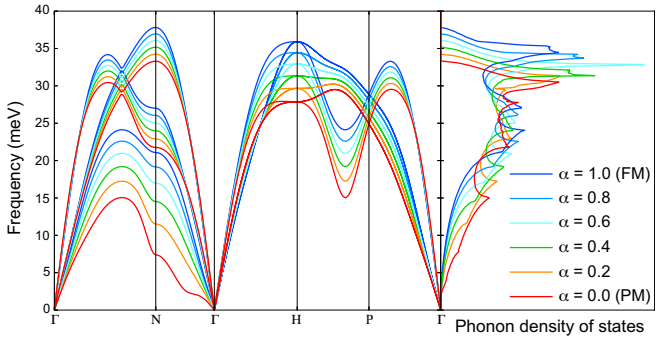} 
\caption{The phonon dispersions and the phonon density of states of bcc Fe from the ferromagnetic state
(FM, $\alpha = 1$) to the paramagnetic state (PM, $\alpha = 0$).
The definition of $\alpha$ is written in Methods.}
\label{phonon_results}
\end{figure}

Phonon frequencies depending on the magnetic energy ($G_{\rm ph}(T, E_{\rm mag})$) can be calculated by using first-principles phonon calculations
and Monte Carlo simulations following the previous research\cite{mag_ph} (see Methods).
The phonon dispersions and the phonon density of states of bcc Fe depending on the magnetic energy are shown in Fig. \ref{phonon_results}.
The dependence of the frequencies on the magnetic energy is represented through the parameter $\alpha$ (see Methods).
The calculated phonon dispersions in the ferromagnetic (FM, $\alpha=1$)
and paramagnetic DLM (PM, $\alpha=0$) limits are consistent
with the previous research\cite{mag_ph}.
Once the phonon frequencies at various magnetic energies (i.e. at various $\alpha$) are calculated, the phonon free energy can be evaluated from the analytical form,
\begin{equation}
G_{\rm ph}(T, E_{\rm mag}) = \frac{k_{\rm B} T}{N_{\bm q}}\sum_{{\bm q},j}\log\left[ 2\sinh\left(\frac{\hbar\omega_{{\bm q}j}(E_{\rm mag})}{2k_{\rm B} T}\right)\right] ,
\label{phonon_F_mag}
\end{equation}
where $k_{\rm B}$ represents the Boltzmann constant, $\omega_{{\bm q}j}(E_{\rm mag})$ the phonon frequency of the $j$-th branch at the wave number $\bm q$ as a function of $E_{\rm mag}$ and $N_{\bm q}$ the total number of $\bm q$ points. 
Note that the more disordered the magnetic state is, the lower the phonon frequencies are.
This tendency means the phonon free energies of paramagnetic states are smaller than that of the ferromagnetic state
because of the monotonicity of the phonon free energy for the phonon frequency.
Consequently, paramagnetic states are thermodynamically stabilised by the phonon softening effect.

\section*{Total free energy minimisation}
We are now able to proceed to the total free energy minimisation in equation (\ref{argmin}) by using the functions $G_{\rm ph}(T, E_{\rm mag})$
and $S_{\rm mag}(E_{\rm mag})$.
The minimisation procedures are simple.
Firstly, we fix the temperature at ${\rm T_0}$.
Secondly, we calculate the total free energy ($G_{\rm ph}({\rm T_0}, E_{\rm mag}) + E_{\rm mag} - {\rm T_0}S_{\rm mag}(E_{\rm mag})$)
for various $E_{\rm mag}$ values. 
The variable range of $E_{\rm mag}$ is from the ferromagnetic limit to the paramagnetic limit.
Thirdly, we find $E_{\rm mag}$ corresponding to the minimum total free energy.
The orange line in Fig. \ref{minimisation} is a visualisation of these steps.
Finally, repeat these steps for a temperature range around $T_{\rm C}$.

\begin{figure}[b]
\includegraphics[width=\linewidth]{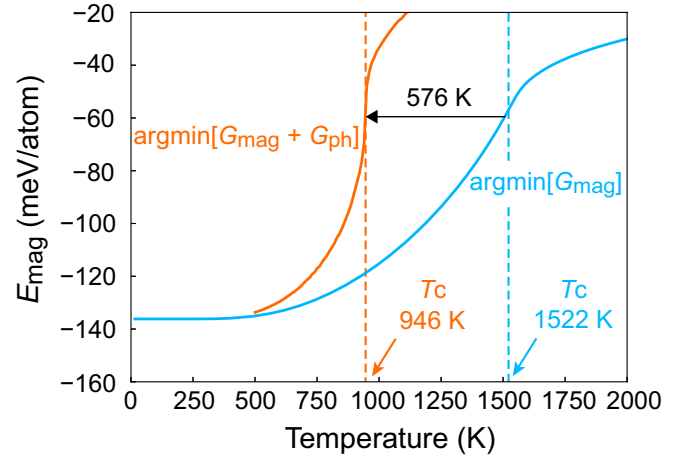} 
\caption{The equilibrium magnetic energy of bcc Fe as a function of temperature.
Orange line represents the equilibrium magnetic energies obtained by the minimisation of the total free energy. 
Blue line represents the equilibrium magnetic energies by the minimisation of the magnetic free energy (the same as the blue line in Fig. \ref{rMC_results} (a)).
The Curie temperature $T_{\rm C}$ in the minimisation of the total free energy is defined as the temperature with the same magnetic energy as the Heisenberg model.
This definition is reasonable because the same magnetic energy gives the same magnetic ordering as long as $J_{ij}$ values do not vary.
}
\label{min_results}
\end{figure}

The equilibrium magnetic energies of bcc Fe obtained by two difference methods are shown in Fig. \ref{min_results}:
One is the minimisation of the total free energy $G_{\rm mag}+G_{\rm ph}$;
the other is the Monte Carlo simulations based on the Heisenberg model (the result is the same as the blue line in Fig. \ref{rMC_results} (a)).
Note that the result from the latter method is corresponding to that of considering only $G_{\rm mag}$ in the minimisation of the free energy.
The equilibrium magnetic energies obtained by the minimisation of the total free energy are larger than
those of considering only $G_{\rm mag}$.
This is, as mentioned before, due to the stabilisation of paramagnetic states by the phonon softening effect, and the magnitude of the stabilisation indicates that the phonon contribution is not negligible at all in the determination of equilibrium magnetic states around $T_{\rm C}$

The stabilisation of paramagnetic states leads to a decrease in $T_{\rm C}$.
As shown in Fig. \ref{min_results}, $T_{\rm C}$ in the results of the minimisation of the total free energy (946 K) is lower than that 
of considering only $G_{\rm mag}$ (1522 K), and the magnitude of the decrease reached nearly 580 K.
Notably, $T_{\rm C}$ of considering both $G_{\rm mag}$ and $G_{\rm ph}$ is dramatically closer to experimental value (1043 K) than
that of considering only $G_{\rm mag}$, i.e. $T_{\rm C}$ in the Heisenberg model.
Although $T_{\rm C}=$ 946 K is still underestimated the experimental value to some extent,
the anharmonicity of phonons probably compensates for the deviation.
Heine, Hellman and Broido\cite{tdep} investigated the phonon softening phenomenon in bcc Fe with including anharmonic effects.
They show that at 1043 K, where the anharmonicity is effective,
the differences between the frequencies of the ferromagnetic and paramagnetic states are reduced compared with those at 300 K.
Thus the difference of the phonon free energies between the ferromagnetic and paramagnetic states is also reduced.
This consequently makes the degree of the decrease in $T_{\rm C}$ smaller than our result.
The underestimation in our result is, therefore, qualitatively correct.

The substantial decrease in $T_{\rm C}$ gives a doubt on the usual recognition of the accuracy in prediction techniques for $T_{\rm C}$.
Roughly speaking, there are three reference magnetic states in the derivation of $J_{ij}$:
ferromagnetic state, paramagnetic DLM state and conical spin-spiral states.
The former two are used within Green's function-based methods\cite{oguchi, mag_force},
whereas conical spin-spiral states are used within the frozen magnon approach\cite{halilov1, halilov2}.
We can summarise the relationship between the reference states and predictive accuracy of $T_{\rm C}$ in bcc Fe:
The ferromagnetic state and spin-spiral states give $T_{\rm C}$ near the experimental
value\cite{mag_force,halilov1,halilov2,rosengaard,pajda,realistic,lezaic},
whereas the paramagnetic DLM state overestimates $T_{\rm C}$ significantly\cite{oguchi,chana,ruban_vib}.
Therefore, the ferromagnetic state and spin-spiral states have been recognised to have an enough predictive accuracy of $T_{\rm C}$ regarding bcc Fe.
However, our study clearly shows that this recognition is questionable
because the contribution of the phonon free energy decreases $T_{\rm C}$ of bcc Fe significantly.
The substantial decrease in $T_{\rm C}$ suggests the DLM state shows correct tendency regarding $T_{\rm C}$ prediction,
rather than the ferromagnetic state and spin-spiral states.
Note that this suggestion is of great importance for theory of finite-temperature magnetism as follows.
In the development of the theory, $T_{\rm C}$ of bcc Fe has been recognised as a touchstone:
Whether the predicted $T_{\rm C}$ of bcc Fe agrees with the experimental value
or not has been an element to examine the validity of a new theory.
Our result, however, indicates such an examination way is inappropriate.
Instead, an appropriate judgment criterion is as follows:
Without considering the phonon softening, a theory that accurately describes finite-temperature magnetism must overestimate $T_{\rm C}$ of bcc Fe.

Our thermodynamic formulation becomes complete if we incorporate the dependence of $E_{\rm mag}$ on $S_{\rm ph}$.
This dependence may be related to the effect of thermal atomic displacements on $J_{ij}$.
Ruban and Peil\cite{ruban_vib} studied this effect by combining $J_{ij}$ calculations and molecular dynamics.
They clearly showed $J_{ij}$ values of bcc Fe were reduced by atomic displacements,
and consequently, $T_{\rm C}$ was also largely decreased compared with the case of excluding thermal atomic displacements.
The dependence of $E_{\rm mag}$ on $S_{\rm ph}$ is thus important and intriguing from a thermodynamic viewpoint.
However, a concrete expression of this dependence is yet to be obtained.


\section*{Conclusions}
We have quantitatively evaluated the thermodynamic feedback effect from phonons to magnetism on $T_{\rm C}$ regarding bcc Fe.
The phonon softening due to magnetic disordering lead to the stabilisation of paramagnetic states.
As a result, $T_{\rm C}$ of bcc Fe was decreased by nearly 580 K from the value in the case of ignoring the feedback effect, i.e. the value for the Heisenberg model.
This deviation in bcc Fe is of great importance because bcc Fe is recognised as a touchstone for the study of finite-temperature magnetism.
We stress two important knowledge regarding the prediction of $T_{\rm C}$:
(i) An appropriate theory of magnetism without considering the contribution of the phonon free energy must overestimate $T_{\rm C}$
in bcc Fe, contrary to conventional understanding.
(ii) We should use $J_{ij}$ for the DLM state rather than for the ferromagnetic state in the accurate description of $T_{\rm C}$.

Finally, we mention the applicability of our thermodynamic formulation.
We focused on bcc Fe in this study, but our formulation is not restricted to it.
It is intriguing to apply the formulation to other magnetic materials such as permanent magnets in which $T_{\rm C}$ is critically important.
In addition, the core concept of the formulation can be applied to other interacting excitation phenomena, not only the interaction between phonons and magnetism:
If a contribution (X) affects other contribution (Y) and changes the free energy of Y,
the thermal equilibrium state of X is also affected through the minimum principle for the free energy.
In our study, X is magnetic states, and Y is phonons.
Therefore, the concept of our thermodynamic formulation can be applied to other interacting excitations
if one can express the magnitude of the interaction as thermodynamic quantities ($E_{\rm mag}$ in our study).
The formulation will be helpful for a quantitative description of the finite-temperature properties of materials.

\clearpage
\section*{methods}
\subsection*{First-principles phonon calculations.} 
All of the phonon calculations were carried out within the harmonic approximation.
To evaluate the phonon frequencies at an intermediate magnetic ordering, we employed a force-averaging method\cite{mag_ph}.
In this method, the atomic forces at an intermediate magnetic ordering are determined by mixing the forces at the ferromagnetic (FM) and paramagnetic (PM) DLM states.
Following the reference\cite{mag_ph}, the atomic forces at an intermediate magnetic ordering can be written as
\begin{equation}
{\bf  F}_i \approx \alpha {\bf F}^{\rm FM}_i + (1-\alpha){\bf F}^{\rm PM}_i,
\end{equation}
where ${\bf  F}_i$ is the atomic force vector on $i$-th atom and $\alpha$ is a mixing parameter.
They also proposed a solid expression of $\alpha$ by using the magnetic energy ($E_{\rm mag}$)  as below:
\begin{equation}
\alpha = \frac{E_{\rm mag} - E_{\rm mag}^{\rm PM}}{E_{\rm mag}^{\rm FM} - E_{\rm mag}^{\rm PM}},
\label{alpha}
\end{equation}
where $E_{\rm mag}^{\rm PM}$ ($E_{\rm mag}^{\rm FM}$) is the magnetic energy at high (low) temperature limit in the Heisenberg model.
In the original paper\cite{mag_ph}, they assumed the temperature dependence of $E_{\rm mag}$ is determined by the Monte Carlo results only.
Therefore, $\alpha$ was treated as a function of temperature ($\alpha = \alpha(\widetilde{T})$).
This is equivalent to that the equilibrium magnetic energy at a temperature is determined to minimise the magnetic free energy, not total free energy.
On the other hand, in our study, $\alpha$ is not regarded as a function of temperature
but is interpreted as a function of energy ($\alpha = \alpha(E)$).
This interpretation allows that the phonon free energy $G_{\rm ph}$ can be
regarded as a function of the magnetic energy ($G_{\rm ph} = G_{\rm ph}(T, E_{\rm mag})$).
The temperature dependence of $E_{\rm mag}$ is determined after the minimisation of the total free energy in equation (\ref{argmin}).
This interpretation is the most important key for solving the minimisation problem in the minimum principle for the free energy.

The paramagnetic DLM state in the phonon calculations was mimicked by a special quasirandom structure\cite{zunger_sqs} on the spin configuration (up and down)
as obtained from the ATAT package\cite{atat}.
The atomic forces were calculated by the direct method\cite{paramag, para_calc}. 
We used the 3$\times$3$\times$3 cubic supercell (54 atoms) for the force calculations in both ferromagnetic and paramagnetic conditions.
The employed lattice constant $a =$ 2.86 {\AA} was derived by combining the relaxed lattice constant and experimental lattice expansion ratio
at $T=1043$ K\cite{lat_exp}.
Although such determination procedure of lattice constant probably gives some pressure even in the framework of the quasiharmonic approximation,
we assume its effect is minor and fixed the volume.
First-principles calculations were based on density functional theory within the projector augmented wave method\cite{PAW},
as implemented in the VASP code\cite{vasp1, vasp2}. For the exchange-correlation functional, the generalised gradient approximation
parametrised  by Perdew, Burke and Ernzerhof\cite{gga} was used.
The cutoff energy 400 eV and 9$\times$9$\times$9 $k$-point grid for the supercell were used for the force calculations.
The derivation of force constants and the calculations of the phonon free energy were performed by using the ALAMODE code\cite{alamode}.

\subsection*{Calculations of exchange coupling constants.}
Exchange coupling constants $J_{ij}$ in the Monte Carlo simulations were derived with magnetic force theorem\cite{mag_force} and
the Korringa-Kohn-Rostoker (KKR) Green's function method along with the coherent potential approximation (CPA)\cite{cpa1,cpa2},
implemented in the AkaiKKR code\cite{cpa2,akaikkr}.
The exchange-correlation functional was treated within the local density approximation\cite{mjw}. 
The lattice constant was set to be the same one as in the phonon calculations.
Paramagnetic DLM state\cite{oguchi,gyorffy} was employed as a reference magnetic state in the derivation of $J_{ij}$.
Calculated $J_{ij}$ values are listed in Table \ref{jij}.

\begin{table}[b]
\caption{Calculated exchange coupling constants $J_{ij}$ of bcc Fe for the paramagnetic DLM state.}
\label{jij}
\begin{center}
\begin{tabularx}{55mm}{X D{.}{.}{-1}} \toprule
nearest neighbour & \multicolumn{1}{c}\texttt{$J_{ij}$ (meV)} \\ \midrule
First & 27.18\\
Second & 2.62\\
Third & 1.33\\
Fourth & 0.21\\
Fifth & -1.37\\ \bottomrule
\end{tabularx}
\end{center}
\end{table}

\subsection*{Monte Carlo simulations.}
To evaluate the magnetic entropy as a function of the magnetic energy, we carried out the classical Monte Carlo simulations
based on the Heisenberg model
\begin{equation}
H = -2\sum_{(i, j)} J_{ij}{\bm e}_i \cdot {\bm e}_j,
\end{equation}
where $J_{ij}$ denotes the exchange coupling constant and ${\bm e}_i$ is the unit vector on site $i$.
We included up to the third nearest neighbour pairs as interacting shells.
The classical Monte Carlo simulations were performed by using the ALPS code\cite{alps}.
To obtain more accurate results,
we employed the rescaled Monte Carlo method\cite{rescaled} which reproduces the quantum specific heat from the classical specific heat. 
The magnetic energy and entropy were derived by integrating the specific heat.
The spin quantum number $S=1.07$ for the DLM condition as calculated by KKR-CPA was used in the rescaled Monte Carlo method.
The Monte Carlo simulations were carried out using a 16$\times$16$\times$16 sites
and involve 300,000 steps for equilibration and 2,700,000 steps for averaging.
Temperature grids of 0.1 and 0.2 mRy were used in the range of near $T_{\rm C}$ and other ranges, respectively.
Note that the entropy in the rescaled Monte Carlo method does not go to zero at $T\rightarrow 0$.
Thus this method is not suitable to describe thermodynamic quantities at a low-temperature range.
Our thermodynamic formulation, however, needs only the result at a temperature range around $T_{\rm C}$.
Thus the shortcoming does not matter in this study.

\subsection*{Author contributions}
The formulation was established by T.T and Y.G.
All of the calculations were conducted by T.T.
The project was supervised by Y.G.
All authors discussed the results and contributed to writing the paper.

\subsection*{Competing interests}
The authors declare no competing interests.

\begin{acknowledgments}
This work was supported in part by MEXT as Fugaku project and the Elements Strategy Initiative Project as well as KAKENHI Grant No. 17K04978.
We are grateful to H. Akai for offering the unreleased version of the AkaiKKR code.
The calculations were partly carried out by using supercomputers at ISSP,
The University of Tokyo, and TSUBAME, Tokyo Institute of Technology as well as the K computer, RIKEN (Project No. hp190169).
\end{acknowledgments}



\end{document}